\documentclass[journal]{IEEEtran}
\usepackage{cite}

\usepackage{booktabs,bbold}

\ifCLASSINFOpdf
   \usepackage[pdftex]{graphicx}
\else
\fi
\usepackage[cmex10]{amsmath}
\usepackage[tight,footnotesize]{subfigure}

\usepackage[caption=false,font=footnotesize]{subfig}

\DeclareMathOperator*{\argmax}{arg\,max}
\providecommand{\abs}[1]{\left|#1\right|}
\begin{document}
%
\title{A Data-Driven Approach to Estimating the Number\\ of Clusters in Hierarchical Clustering*}
%
%
%

\author{Antoine~E.~Zambelli\thanks{*This is a preprint.} %
	}

%
%

\markboth{Journal of \LaTeX\ Class Files,~Vol.~6, No.~1, January~2007}%
{Shell \MakeLowercase{\textit{et al.}}: Bare Demo of IEEEtran.cls for Journals}
%



\maketitle
\thispagestyle{empty}

\begin{abstract}
We propose two new methods for estimating the number of clusters in a hierarchical clustering framework in the hopes of creating a fully automated process with no human intervention. The methods are completely data-driven and require no input from the researcher, and as such are fully automated. They are quite easy to implement and not computationally intensive in the least. We analyze performance on several simulated data sets and the Biobase Gene Expression Set, comparing our methods to the established Gap statistic and Elbow methods and outperforming both in multi-cluster scenarios.
\end{abstract}


%

\section{Introduction}
%
%
%
%
\IEEEPARstart{H}{ierarchical} clustering analysis (or HCA) is an extensively studied field of unsupervised learning. Very useful in dimensionality reduction problems, we will study ways of using this clustering method with the aim of reducing (or removing) the need for human intervention.

This problem of human intervention stems from the fact that HCA is used when we do not know the correct number of clusters in our data (otherwise we might use, say, K-means). While the ability to cluster data with an unknown number of clusters is a powerful one, we often need a researcher to interpret the results - or cutoff the algorithm - to recover a meaningful cluster number.

While our motivation stems from DNA micro-array data and Gene Expression problems, these methods can apply to any similarly structured scenario. Specifically, we will analyze different existing automated methods for cutting off HCA and propose two new ones.

In Section II we will discuss background material on HCA and the existing methods and in Section III we will present some technical details on these methods and introduce our own. Section 4 will contain results on simulated and actual data, and Section 5 will examine data sampling procedures to improve accuracy.

\section{Background}

Hierarchical clustering, briefly, seeks to pair up data points that are most similar to one another. With the agglomerative (or bottom-up) approach, we begin with $N$ data points forming singleton clusters. For each point, we measure the distance between it and its $N-1$ neighbors. The pair with the shortest distance between them is taken to form a new cluster. We then look at the distance between the $N-2$ points remaining and the newly formed cluster, and again pair off the two with shortest distance (either adding to our 2-cluster, or forming another one). This process is repeated until we have a single cluster with $N$ points (regardless of the absolute distance between points).

Naturally, this is a very good dimensionality reduction algorithm. Unfortunately, it keeps going until we've flattened our data to 1 dimension. In cases where in truth we have $n\geq 2$ clusters, this is problematic.

\begin{figure}[h!]
	\centering\includegraphics[height=5cm]{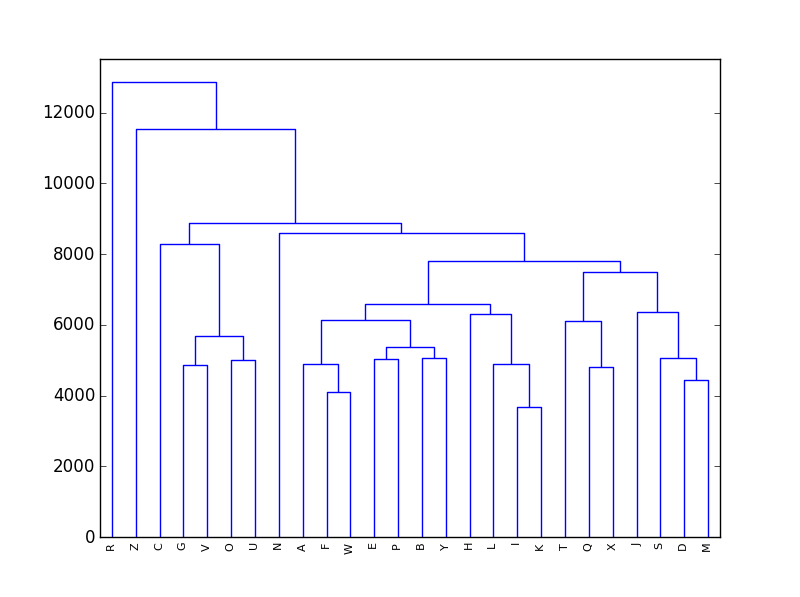}
	\caption{Example dendrogram from 26 data points.}\label{fig:1}
\end{figure}

The results of a HCA are often expressed as a dendrogram, a tree-like graph that contains vital information about the distances measured in the clustering and the pairings generated. An example of a dendrogram can be seen in Figure~\ref{fig:1}.
Briefly, horizontal lines denote pairings, and the height of those lines represent the distance that needed to be bridged in order to cluster the points together. That is, the smaller the height (or jump) of a pairing, the closer the points were to begin with.

Our goal is to find a way to say, once the full tree is made, ``jumps beyond this point are not reasonable", and we can cutoff the algorithm, keeping only those clusters generated before that point.

\subsection{Existing Alternatives}

The problem of cutting off a dendrogram is one that researchers encounter often, but there are no reliable automated methods for doing it. Often, the Gap statistic is the only proposed automated method, as in~\cite{hastie}. As such, many researchers will inspect the finished dendrogram and manually select a cutoff point, based on their own judgment. Apart from the obviously slow nature of this exercise, there is also the question of human error to consider - as well as bias. In cases where the cutoff is not easily determined, two different researchers may arrive at different conclusions as to the correct number of clusters - which could both be incorrect. Algorithmic approaches aim to eliminate this, and range from simpler methods to more complex ones. An excellent summary of existing methods is given in~\cite{gordon}, which is in fact referenced in~\cite{tib}.

The latter, more importantly, develops the Gap statistic. We will present the technical aspects in Section III, but we quickly discuss some properties here. First, the Gap statistic is one of few methods that is capable of accurately estimating single clusters (in the case where all our data belongs to one cluster), a situation often undefined for other methods. While it is rather precise overall, it requires the use of a ``reference distribution", which must be chosen by the researcher. They put forward that the uniform distribution is in fact the best choice for unimodal distributions. A powerful result, it is still limited in other cases, and thus many researchers still take the manual approach. However, it generally outperforms other complex methods, as such we focus on the Gap statistic.

On the other side of the complexity spectrum, we have variants of the ``Elbow method". The elbow method looks to explain the variance in the data as a function of the number of clusters we assign. The more clusters we assign, the more variance we can explain. However, the marginal gain from adding new clusters will begin to diminish - we choose this point as the number of clusters. A variant of this method, often applied to dendrograms, looks for the largest acceleration of distance growth~\cite{hees}. While this method is very flexible, it cannot handle the single-cluster case.

\section{Approaches}

We will look at both the elbow method variant and the Gap statistic, as well as our own 2 methods we are presenting in this paper. While there are many other methods to compare them to, the Gap statistic is quite representative of a successful (if more complex) solution - and tends to outperform the other known methods. The elbow method represents the more accepted simple approaches. In all tests in this paper, we use an agglomerative hierarchy, with average linkage and euclidean distance measure.

\subsection{Gap Statistic}

The Gap statistic is constructed from the within-cluster distances, and comparing their sum to the expected value under a null distribution. Specifically, as given in~\cite{tib}, we have for $r$ clusters $C_r$
\begin{equation}
Gap_n(k) = E^*_n[\log W_k] - \log W_k
\end{equation}
where, with $n_r = \abs{C_r}$,
\begin{equation}
W_k = \sum_{r=1}^k{\frac{1}{2n_r}D_r} = \sum_{r=1}^k{\frac{1}{2n_r}\sum_{i,i'\in C_r}{d_{ii'}}}
\end{equation}
That is, we are looking at the sum of the within-cluster distances $d$, across all $r$ clusters $C_r$. Computationally, we estimate the Gap statistic and find the number of clusters to be (as per~\cite{tib})
\begin{equation}
\hat{k}_G = \textrm{smallest $k\mid$ $Gap(k)\geq Gap(k+1)-s_{k+1}$}
\end{equation}
where $s_k$ is the standard error from the estimation of $Gap(k)$. As mentioned,~\cite{tib} considers both a uniform distribution approach and a principal component construction. In many cases, the uniform distribution performs better, and this is the one we will use.

\subsection{Elbow Method}

This variant of the elbow method, which looks at the acceleration, is seen in~\cite{hees}. A straightforward method, we simply look at the acceleration in jump sizes. So given the set of distances from our clustering $\{d_1,\cdots,d_N\}$, the acceleration can be written as
\begin{equation}
\left\{ d_3-2d_2-d_1,\cdots,d_N-2d_{N-1}-d_{N-2} \right\}.
\end{equation}
We choose our number of clusters as the jump with the highest acceleration, giving us
\begin{equation}
\hat{k}_E = N+2 - \argmax_{i\in[3,N]} \left\{ d_i-2d_{i-1}-d_{i-2} \right\}.
\end{equation}
While very simple and very fast, this method will never find the endpoints, ie, the $N$ singleton clusters and the single $N$-element cluster cases.

\subsection{Mode}

The first method we propose is based on the empirical distribution of jump sizes. Specifically, we look to the mode of the distribution $D=\{d_1,\cdots,d_N\}$, denoted $\hat{D}$, adjusted by the standard deviation ($\sigma_D$). Our motivation is that the most common jump size likely does not represent a good cutoff point, and we should look at a higher jump threshold. As such, we take the number of clusters to be
\begin{equation}
\hat{k}_M = \hat{D} + \alpha\sigma_D,
\end{equation}
where $\alpha$ is a parameter that can be tuned. Naturally, tuning $\alpha$ would require human intervention or a training data set. As such, we set it to $3$ somewhat arbitrarily.

\subsection{Maximum Difference}

Our second method is even simpler, but is surprisingly absent from the literature. Inspired by the elbow method, we look at the maximum jump difference - as opposed to acceleration. Our number of clusters is then given by
\begin{equation}
\hat{k}_D = N+2 - \argmax_{i\in[2,N]}\left\{ d_i-d_{i-1} \right\}.
\end{equation}
This method shares the elbow method's drawback that it cannot solve the single cluster case (though it can handle singleton clusters), but we thought it prudent to examine as the literature seemed to focus on acceleration and not velocity.

\section{Results}

We present results of simulations on several numbers of true clusters, drawn from a $2$-dimensional normal distribution. Each cluster is comprised of $100$ samples. We are most interested in tracking the success rate and the \emph{error size given an incorrect estimate}. That is, how often can we correctly estimate the number of clusters $k$ and when we can't, by how much are we off? Formally, this is given by
\begin{equation}
S_X = \frac{\abs{T}}{n},\ \textrm{with} \ T = \left\{ j \mid \hat{k}_X^{(j)} = k  \right\}_{j=1}^n
\end{equation}
and
\begin{equation}
E_X = \frac{1}{\abs{S}} \sum_{x\in S}{ x },\ \textrm{with} \ S=\left\{ \abs{k-\hat{k}_X^{(j)}}\mid k\neq \hat{k}_X^{(j)} \right\}_{j=1}^n.
\end{equation}
We chose this measure of the error to avoid under-estimating error size (which would happen in a method that is often correct).

\subsection{Simulated Data}

\begin{figure}[h!]
	\centering\includegraphics[height=5cm]{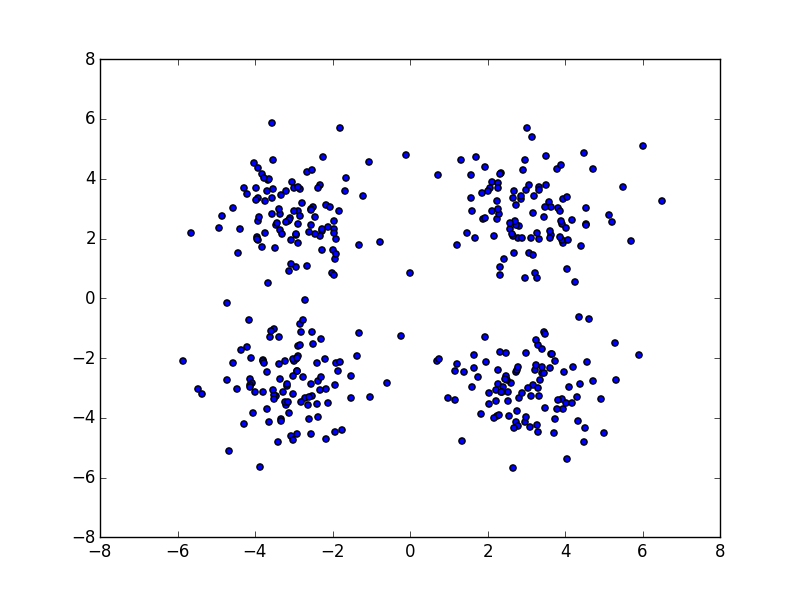}
	\caption{Sample clusters of $100$ points each.}\label{fig:2}
\end{figure}

The data used was drawn from a standard normal distribution, with cluster centers at $(-3,-3),(3,3),(-3,3),(3,-3)$, shown in Figure~\ref{fig:2}. In the case of $1$ cluster, the first is taken, for $2$ clusters, the first two, and so on. We present the results of the methods on $n=200$ simulations below in Tables I-III, with the best results in bold.

\begin{table}[h!]
	\caption{Average cluster numbers over $n=200$ runs.}
	\centering\begin{tabular}{l|rrrr}
		\toprule
		$k$ &  $\bar{\hat{k}}_E$ & $\bar{\hat{k}}_G$ &  $\bar{\hat{k}}_D$ &  $\bar{\hat{k}}_M$ \\
		\midrule
		1 &  3.720 &  \textbf{1.110} &     2.965 &     4.505 \\
		2 &  \textbf{2.000} &  \textbf{2.000} &     \textbf{2.000} &     2.645 \\
		3 &  3.050 &  3.050 &     \textbf{3.015} &     4.725 \\
		4 &  3.915 &  4.060 &     \textbf{4.010} &     6.280 \\
		\bottomrule
	\end{tabular}
\end{table}
\begin{table}[h!]
	\caption{Success rate over $n=200$ runs.}
	\centering\begin{tabular}{l|rrrr}
		\toprule
		$k$ &  $S_E$ & $S_G$ &  $S_D$ &  $S_M$ \\
		\midrule
		1 &  0.000 &  \textbf{0.910} &     0.000 &     0.000 \\
		2 &  \textbf{1.000} &  \textbf{1.000} &     \textbf{1.000} &     0.465 \\
		3 &  0.955 &  0.960 &     \textbf{0.985} &     0.100 \\
		4 &  0.920 &  0.940 &     \textbf{0.990} &     0.055 \\
		\bottomrule
	\end{tabular}
\end{table}
\begin{table}[h!]
	\caption{Average error size (when wrong) over $n=200$ runs.}
	\centering\begin{tabular}{l|rrrr}
		\toprule
		$k$ &  $E_E$ & $E_G$ &  $E_D$ &  $E_M$ \\
		\midrule
		1 &  2.720 &  \textbf{1.222} &     1.965 &  3.505 \\
		2 &  \textbf{0.000} &  \textbf{0.000} &     \textbf{0.000} &  1.206 \\
		3 &  1.111 &  1.250 &     \textbf{1.000} &  1.917 \\
		4 &  1.688 &  \textbf{1.000} &     \textbf{1.000} &  2.413 \\
		\bottomrule
	\end{tabular}
\end{table}
As we can see, the best performing methods are so far the gap statistic and, surprisingly, the maximum difference. The maximum difference method has a near perfect success rate on this simple example, besting the Gap statistic in most areas. As noted though, it suffers from the same problem as the elbow method in that it cannot handle the single-cluster case. It is our recommendation that if the reader suspects their data may be a single cluster, they should consider the gap statistic method. Note however that it is much more computationally intensive (by a factor of $\sim 50000$).

\subsection{Different Cluster Distributions}

To get a better sense for the behavior of these methods, we will look at clusters drawn from normal distributions with different parameters. For the clusters centered at $(-3,-3),(3,3),(-3,3),(3,-3)$, we scale the standard deviations to use, respectively: $\mathbb{1}_2,\mathbb{1}_2,2\mathbb{1}_2,0.5\mathbb{1}_2$.
Hopefully, this can show us some limitations of these methods. the results are detailed in Tables IV-VI.

\begin{table}[h!]
	\caption{Average cluster numbers over $n=200$ runs.}
	\centering\begin{tabular}{l|rrrr}
		\toprule
		$k$ & $\bar{\hat{k}}_E$ &  $\bar{\hat{k}}_G$ &  $\bar{\hat{k}}_D$ &  $\bar{\hat{k}}_M$ \\
		\midrule
		1 &   3.710 &  \textbf{1.100} &     3.035 &     4.450 \\
		2 &   \textbf{2.000} &  \textbf{2.000} &     \textbf{2.000} &     2.625 \\
		3 &   3.170 &  \textbf{3.000} &     3.060 &     6.055 \\
		4 &   3.940 &  4.140 &     \textbf{4.005} &     5.780 \\
		\bottomrule
	\end{tabular}
\end{table}
\begin{table}[h!]
	\caption{Success rate over $n=200$ runs.}
	\centering\begin{tabular}{l|rrrr}
		\toprule
		$k$ & $S_E$ &  $S_G$ &  $S_D$ &  $S_M$ \\
		\midrule
		1 &  0.000 &  \textbf{0.910} &     0.000 &     0.000 \\
		2 &  \textbf{1.000} &  \textbf{1.000} &     \textbf{1.000} &     0.475 \\
		3 &  0.840 &  \textbf{1.000} &     0.920 &     0.005 \\
		4 &  0.955 &  0.880 &     \textbf{0.995} &     0.085 \\
		\bottomrule
	\end{tabular}
\end{table}
\begin{table}[h!]
	\caption{Average error size (when wrong) over $n=200$ runs.}
	\centering\begin{tabular}{l|rrrr}
		\toprule
		$k$ & $E_E$ &  $E_G$ &  $E_D$ &  $E_M$ \\
		\midrule
		1 &  2.710 &  \textbf{1.111} &     2.035 &  3.450 \\
		2 &  \textbf{0.000} &  \textbf{0.000} &     \textbf{0.000} &  1.190 \\
		3 &  1.063 &  \textbf{0.000} &     1.000 &  3.070 \\
		4 &  1.778 &  1.167 &     \textbf{1.000} &  1.945 \\
		\bottomrule
	\end{tabular}
\end{table}
In this more complicated case, we can see similar results. The $3$-cluster case (in this example) seems problematic for our method and the elbow method. The Gap statistic once again performs well for the single-cluster scenario, but shows some weakness when we get to $4$ clusters. Overall, it seems that for $k=2,3,4$ clusters, the maximum difference method at the very least keeps up with the Gap statistic, and improves on it in certain cases. 

\subsection{Different Cluster Separations}

\begin{figure}[h!]
	\centering\includegraphics[width=\columnwidth]{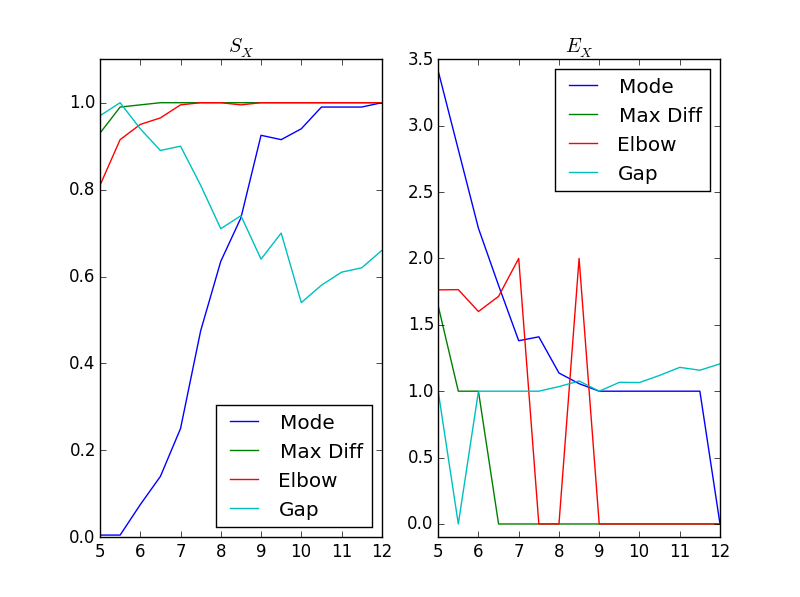}
	\caption{Tracking for varying distances.}\label{fig:3}
\end{figure}

Returning to the equal-distribution $4$-cluster problem, we now look at how our metrics evolve as we increase the distance between the clusters. On the $x$-axis in Figure~\ref{fig:3}, a value of $m$ corresponds to a cluster arrangement with coordinates: $(-m/2,-m/2),(m/2,m/2),(-m/2,m/2),(m/2,-m/2)$. We would expect all methods to perform better as the clusters drift further apart, since they are then more distinct.

This is indeed the case for the elbow, maximum difference and mode methods, which converge to a success rate of $1$ and an error size of $0$ (note some increased stability in the maximum difference). However, the Gap statistic appears to do worse as the clusters separate - which could point to some underlying issues and should be explored more fully.

\subsection{Biobase Set}

\begin{figure}
	\centering\includegraphics[width=\columnwidth]{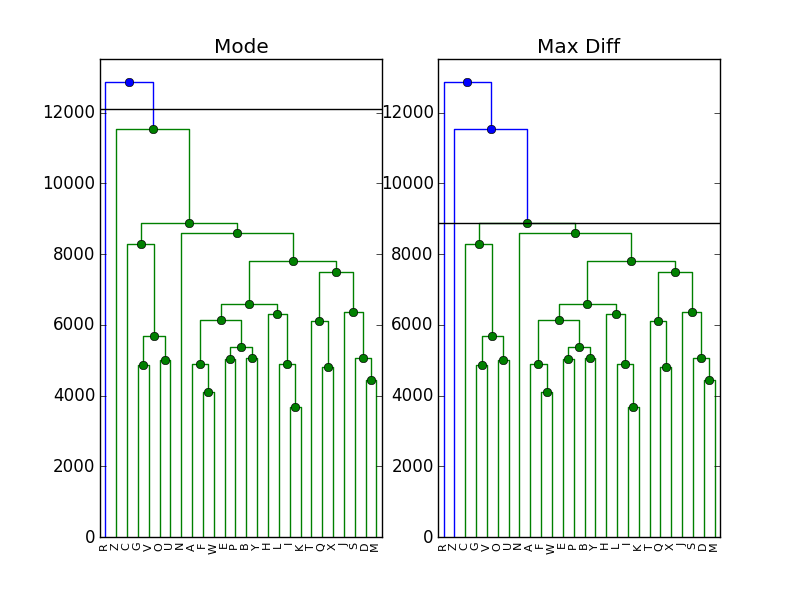}
	\caption{Clustering of Biobase Expression Set.}\label{fig:4}
\end{figure}

\begin{figure}
	\centering\includegraphics[width=\columnwidth]{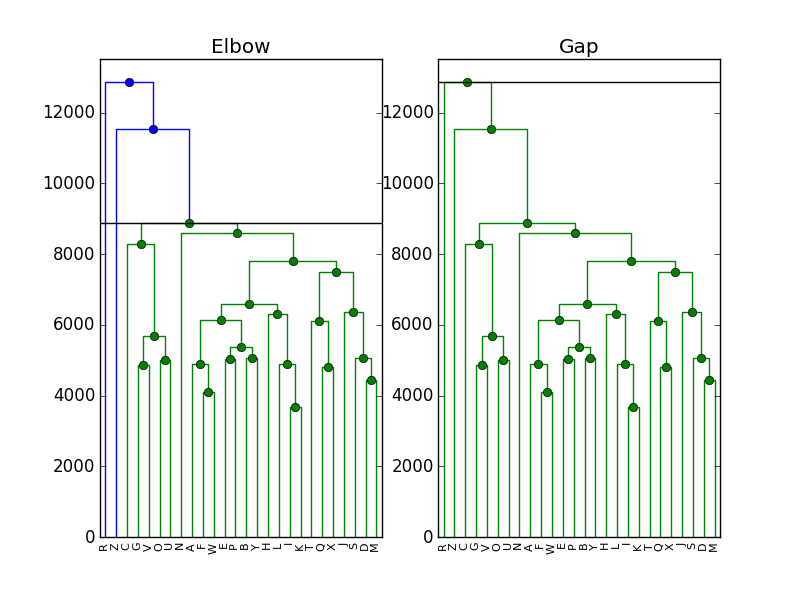}
	\caption{Clustering of Biobase Expression Set.}\label{fig:5}
\end{figure}

As we mentioned back Section I, our primary motivation for this problem was that of DNA microarray data and gene expression problems. It is also always prudent to test things out on real data. As such (and to help with reproducibility), we will test our methods on the Expression Set data from the R package Biobase (Bioconductor),~\cite{huber}. This is a sample of $26$ different elements, with reconstructions presented in Figures~\ref{fig:4} and~\ref{fig:5}.

In this case, the maximum difference and elbow methods were in agreement and selected $\hat{k}_E=\hat{k}_D=3$ clusters (with samples $R$ and $Z$ being singleton clusters). The mode, however, chose to add $Z$ into the main cluster, producing $\hat{k}_M = 2$ final clusters. on the other hand, the Gap statistic selects only $\hat{k}_G=1$ cluster.

The author finds both the mode and Gap results to be somewhat dubious - but they highlight an important issue. How can we know that a clustering is correct? Even if we examine the dendrogram as we did here, it is likely that in many examples the cutoff point could be debated. In this data set, we find it more challenging to determine a correct clustering between $2$ and $3$ clusters - though $3$ seems more natural to the author. This calls back to the previously mentioned issue with manual cutoff selection.

\section{Data Mixing}

In an effort to improve our new methods, we look at data sampling. Inspired by Cross-Validation methods, we will randomly sample $M=\frac{N}{2}$ points $L=100$ times. For each of the $L$ samples $j$, we then run our method and get a $\hat{k}_X^{(j)}$. We then set our estimated number of clusters to be
\begin{equation}
\hat{k}_X^* = \textrm{mode}\left\{ \hat{k}_X^{(j)} \right\}_{j=1}^L.
\end{equation}
While this requires running our method $L$ times, for $L=100$, we are still roughly $500$ times faster that the Gap statistic. Hopefully, this will improve our methods by averaging out any outlying errors in our methods or points in our data.

\subsection{Results}

We present results on the same $k=2,3,4$ cluster construction detailed above, each with the same distribution. Due to computational times, we did not do data mixing on the Gap statistic. Though given that these methods are so much faster even with mixing, we believe that comparing them remains a fruitful exercise. We provide the Gap statistic results from Tables I-III here for convenience (note that no mixing was done for the Gap statistic).

\begin{table}[h!]
	\caption{Average cluster numbers over $n=200$ runs.}
	\centering\begin{tabular}{l|rrrr}
		\toprule
		$k$ & $\hat{k}_E^*$ & $\bar{\hat{k}}_G$ &  $\hat{k}_D^*$ &  $\hat{k}_M^*$ \\
		\midrule
		2 &      \textbf{2.000} & \textbf{2.000}  &      \textbf{2.000} &     2.015 \\
		3 &      \textbf{3.000} & 3.050  &      \textbf{3.000} &     3.080 \\
		4 &      \textbf{4.000} & 4.060  &      \textbf{4.000} &     4.065 \\
		\bottomrule
	\end{tabular}
\end{table}
\begin{table}[h!]
	\caption{Success rate over $n=200$ runs.}
	\centering\begin{tabular}{l|rrrr}
		\toprule
		$k$ & $S_E^*$ & $S_G$ &  $S_D^*$ &  $S_M^*$ \\
		\midrule
		2 &      \textbf{1.000} &  \textbf{1.000}  &     \textbf{1.000} &     0.985 \\
		3 &      \textbf{1.000} &  0.960  &     \textbf{1.000} &     0.920 \\
		4 &      \textbf{1.000} &  0.940  &     \textbf{1.000} &     0.935 \\
		\bottomrule
	\end{tabular}
\end{table}
\begin{table}[h!]
	\caption{Average error size (when wrong) over $n=200$ runs.}
	\centering\begin{tabular}{l|rrrr}
		\toprule
		$k$ & $E_E^*$ & $E_G$ &  $E_D^*$ &  $E_M^*$ \\
		\midrule
		2 &      \textbf{0.000} &  \textbf{0.000}  &     \textbf{0.000} &         1.000 \\
		3 &      \textbf{0.000} &  1.250  &     \textbf{0.000} &         1.000 \\
		4 &      \textbf{0.000} &  1.000  &     \textbf{0.000} &         1.000 \\
		\bottomrule
	\end{tabular}
\end{table}
As we can see, both the elbow method and the maximum difference seem to perfectly capture our simulated data. Perhaps even more surprising, the mode method now has results that are comparable to the Gap statistic and not far behind our other methods.

\subsection{Biobase Set with Mixing}

\begin{figure}
	\centering\includegraphics[width=\columnwidth]{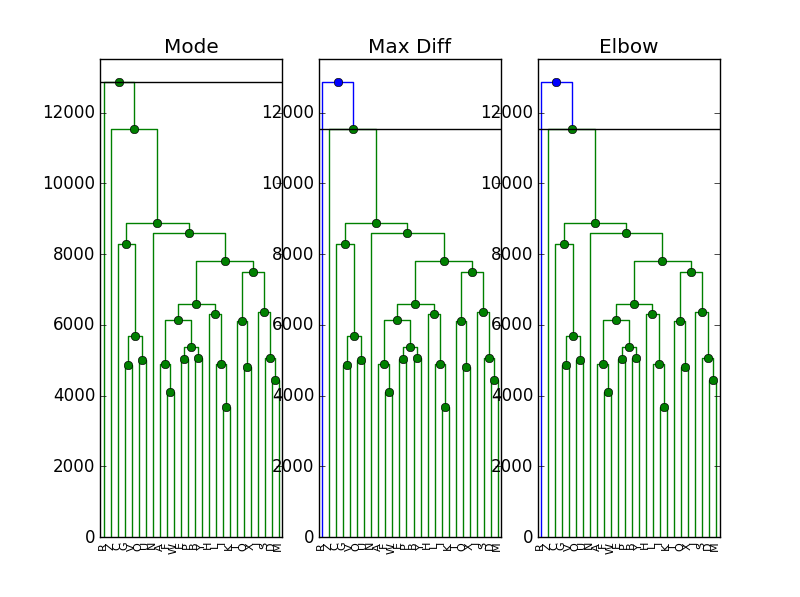}
	\caption{Clustering of Biobase Expression Set with Mixing.}\label{fig:6}
\end{figure}

We now return to our Biobase Expression Set to see if we can come to a consensus on its clustering. We run the same $M=\frac{N}{2}$ and $L=100$ mixing as for the simulated data to obtain Figure~\ref{fig:6}.
With mixing, we get slight differences in our clustering. The elbow and maximum difference methods now agree on $2$ clusters instead of $3$. The mode method agrees with the Gap statistic and sets the number of clusters to $1$. Again, we find it difficult to argue in favor of $1$ cluster, but maintain that $2$ or $3$ clusters seem viable - with a preference for $3$. It is possible we somehow over-mixed the data when working with such a small sample.
\newpage
\subsection{LOOCV}

Our data mixing procedure begins to resemble Leave $N-M$ out Cross Validation. So let's extend this to resemble Leave One Out Cross-Validation (note that computation times will now increase with data size). In our case, this means taking $M=1$ and removing each data point once (in a sense, $L=N$). For each ``sampled" set (of $N-1$ points), we compute the number of clusters and again take the mode of the $L$ sample cluster numbers as our estimate. With this we obtain Tables X-XII.

\begin{table}[h!]
	\caption{Average cluster numbers over $n=200$ runs.}
	\centering\begin{tabular}{l|rrrr}
		\toprule
		$k$ & $\hat{k}_E^*$ & $\bar{\hat{k}}_G$ &  $\hat{k}_D^*$ &  $\hat{k}_M^*$ \\
		\midrule
		2 &   \textbf{2.000} & \textbf{2.000}   &     \textbf{2.000} &      2.660 \\
		3 &   3.040 & 3.050  &     \textbf{3.000} &      4.630 \\
		4 &   3.900 & 4.060   &     \textbf{4.000} &      6.180 \\
		\bottomrule
	\end{tabular}
\end{table}
\begin{table}[h!]
	\caption{Success rate over $n=200$ runs.}
	\centering\begin{tabular}{l|rrrr}
		\toprule
		$k$ & $S_E^*$ & $S_G$ &  $S_D^*$ &  $S_M^*$ \\
		\midrule
		2 &   \textbf{1.000} &  \textbf{1.000}  &     \textbf{1.000} &     0.465 \\
		3 &   0.960 &  0.960  &     \textbf{1.000} &     0.135 \\
		4 &   0.940 &  0.940  &     \textbf{1.000} &     0.055 \\
		\bottomrule
	\end{tabular}
\end{table}
\begin{table}[h!]
	\caption{Average error size (when wrong) over $n=200$ runs.}
	\centering\begin{tabular}{l|rrrr}
		\toprule
		$k$ & $E_E^*$ & $E_G$ &  $E_D^*$ &  $E_M^*$ \\
		\midrule
		2 &      \textbf{0.000} &  \textbf{0.000}  &     \textbf{0.000} &  1.234 \\
		3 &      1.000 &  1.250  &     \textbf{0.000} &  1.884 \\
		4 &      2.000 &  1.000  &     \textbf{0.000} &  2.307 \\
		\bottomrule
	\end{tabular}
\end{table}
While the maximum difference method seems robust in the face of different sampling, it seems that this exercise has revealed some instability in the mode method, which has reverted back to a lackluster performance. To a much lesser extent, the elbow method is having some more trouble as well.

It seems more likely that the choice of sampling parameters could be the cause of the clustering in the Biobase data in Figure~\ref{fig:6}. More generally, we should look into determining optimal mixing parameters $M$ and $L$ and/or their impact on these methods.

\begin{figure}[h!]
	\centering\includegraphics[width=\columnwidth]{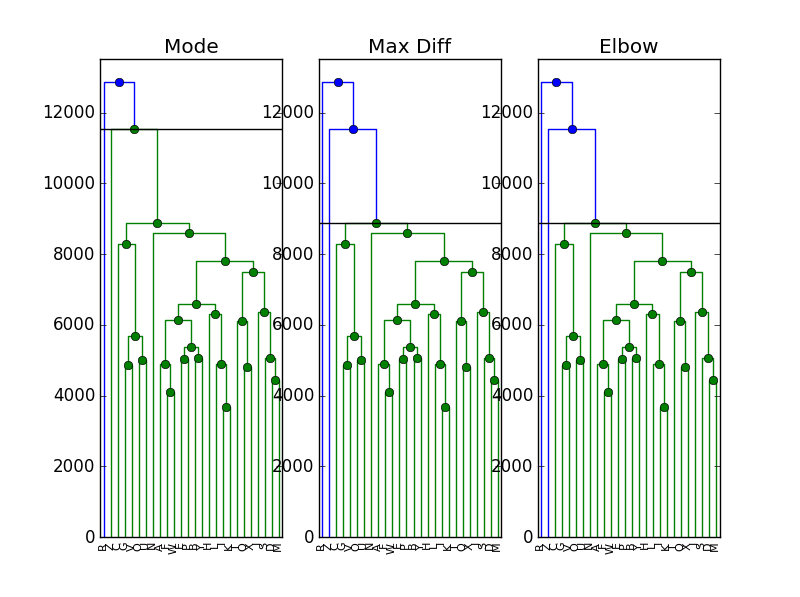}
	\caption{Clustering of Biobase Expression Set with ``LOOCV".}\label{fig:7}
\end{figure}

This mixing method does appear to perform better for the Gene Expression set than the previous choice of mixing parameters, which seems to confirm our hypothesis that there is perhaps an oversampling effect that can come into play, or something along those lines which much be explored more fully.

\section{Conclusion}

We have developed two new empirical methods for clustering data in a hierarchical framework. While our methods are substantially faster than the existing Gap statistic, we insist that they do not handle the single-cluster case. In other cases, our maximum difference method is at least comparable to the Gap statistic and outperforms the elbow method.

In addition, the use of the data mixing procedure presented can greatly improve performance (especially for the mode method), leading to the maximum difference method outperforming the other 3. Lastly, these methods can be implemented in a few lines of code and should allow researchers to quickly utilize them at little computational cost.

In the future we hope to study the possibility of finding optimal mixing numbers $M$ and $L$ and the impact of the choice of these parameters of our results. Hopefully, they are related to the instability detected in the mode method when using $M=1$ in our mixing procedure.

\ifCLASSOPTIONcaptionsoff
  \newpage
\fi



\bibliographystyle{IEEEtran}
%

%
\newpage






\end{document}